\documentclass[nofootinbib,aps,prab,reprint,groupedaddress,showpacs]{revtex4-2}
\usepackage[english]{babel}

\usepackage[utf8x]{inputenc}
\usepackage{amssymb}
\usepackage{amsthm}
\usepackage{amsmath}
\DeclareMathOperator\arctanh{arctanh}
\usepackage{hyperref}
\usepackage{bm}

\hypersetup{
    colorlinks=true, 
    linkcolor=cyan,
    citecolor=magenta, 
    filecolor=magenta, 
    urlcolor=cyan,
    runcolor=cyan
}

\usepackage{subfigure}
\usepackage[percent]{overpic} 
\usepackage{picture}
\usepackage{color,soul}
\usepackage{nth}
\usepackage{gensymb}
\usepackage{cleveref}
\usepackage{textcomp}
\usepackage{makecell}
\usepackage{lineno} 
\usepackage{textcomp}
\usepackage{siunitx}

\graphicspath{{img/}}
\usepackage{comment}

\usepackage{bm}
\usepackage{physics}

\usepackage[english]{babel}

\makeatletter
\renewcommand{\fnum@figure}{FIG. \thefigure}
\makeatother

\begin{document}

\title{Extreme radiation emission regime for electron beams in strong focusing ion channels and undulators}

\author{A. Frazzitta$^{1,2,*}$, M. Yadav$^{3,*}$, J. Mann$^{3}$, A.R. Rossi$^{2}$ and J. B. Rosenzweig$^{3}$}

\affiliation{
$^{1}$Department of Physics, University of Rome “La Sapienza”, p.le A. Moro, 2-00185 Rome, Italy\\
$^{2}$Department of Physics, INFN—Milan, Via Celoria, 16-20133 Milan, Italy\\
$^{3}$Department of Physics and Astronomy, University of California, Los Angeles, USA
}

\email{andrea.frazzitta@uniroma1.it, yadavmonika@g.ucla.edu}
\thanks{These two authors contributed equally}

\date{\today}

\begin{abstract}
A fundamental comparison between undulator and ion channel radiation is presented. Conventional theory for both devices fails to describe high $k$ and $K/\gamma$ regimes accurately, providing an underestimation of particle trajectory amplitude and period. This may lead to incorrect estimation of radiation emission in many setups of practical interest, such as the ion column. A redefinition of plasma density and undulator strength expressions leads to a more reliable prediction of particle behaviour, reproducing the closest possible conditions in the two devices and correctly matching expected betatron oscillation amplitude and wavelength for a wide range of $K/\gamma$ values. Differences in spectral features of the two devices can then be addressed via numerical simulations of single particle and beam dynamics. In this paper we outline a theoretical framework and compare its results with numerical simulation applied to setups eligible for possible radiation sources.
\end{abstract}

\maketitle

\section{\label{sec:introduction} Introduction}

Understanding the radiation mechanisms in plasma channels and undulators is crucial for developing compact, high-brightness X-ray sources and advancing accelerator concepts \cite{Esarey:2002rp,TajimaDawson1979,Dawson1959NonlinearPlasma,esarey1996overview}. For example, FACET-II \cite{facet2}, FLASHForward \cite{Schroder2020TunableFLASHForward}, and the EuPRAXIA \text{@}SPARC\_LAB facility at INFN Frascati integrate advanced electron and laser beam technologies to explore plasma acceleration, THz radiation, and high-brightness electron beams \cite{FERRARIO_2013,Ricardo_2024}. In undulators, charged particles undergo oscillations due to the periodic magnetic field generated by the undulator’s structure. This configuration is essential for synchrotron radiation facilities and free-electron lasers, enabling the generation of high-brightness, coherent radiation \cite{Hidding_PRL_2012,ManahanDechirper2017}. In plasma channels, when electrons are injected into the ion channel (IC), they experience transverse oscillations due to the ion channel’s focusing forces \cite{jamieblowout,ionmotion_rosenzweig_2005}. Contrary to an ideal undulator, in an IC the field is electric and linearly focuses particles towards the device's axis, leading to betatron oscillations and betatron radiation \cite{yadav1,yadav2,Andrea-Frazzitta}. The period and strength of these oscillations are determined, in linear approximation, by the initial conditions of the electrons and the plasma wakefield characteristics \cite{Viktor_1989, Wilks_1989_photon_acc,Mori_1997,Jamie_1997,Joshi2018PlasmaII}. In the present paper, an analytical description for more extreme nonlinear regimes will be presented. 

For large IC particle oscillations ($(K\gg1$), the wiggler regime) the spectrum contains multiple harmonics up to the critical frequency, and for small oscillations ($(K\leq1$), the undulator regime) the spectrum is nearly monochromatic. Up to first order, radiation is emitted within a narrow cone, centered on the device's axis, of angle $\theta = 1/\gamma_0$ or $\theta = K/\gamma_0$ for the undulator and wiggler regimes, respectively \cite{Corde_2013,SakaiICS2017}. The highest frequency in the radiation spectrum, determined by the electron's trajectory and the radius of curvature $\rho$, is defined by $\omega_c$ and should be found at \(\theta=0\). We will see that this behavior is violated in the high $(K/\gamma_0)$ regime.

Betatron radiation can be analyzed using the principles of moving charge radiation~\cite{Jackson1998}. The ion cavity acts as an undulator or a wiggler, characterized by a period $\lambda_u(t)$ and a strength parameter $K(t)$, depending on both the plasma density and the electron's initial conditions upon entering the cavity. Betatron radiation differs from undulator radiation due to its dependence on the beam's initial distance from the axis and the corresponding oscillation of the beam energy, which in extreme regimes can double the beam energy or more. The spectrum characteristics are well-defined in cases of zero longitudinal acceleration and low oscillation strength, but a comprehensive analytical model for general cases is lacking. We will show radiation properties as a function of the energy oscillation amplitude. We compare betatron and undulator radiation in high $(K/\gamma_0$) regime, focusing on divergence and critical frequency definitions. The conditions for an approximate matching of betatron wavelength and oscillation amplitude in the two devices will be presented.

The paper is organized as follows. In Sec.~\ref{gen_wig_traj}, we discuss a correction to the definition of undulator strength parameter $(K$), introducing fully analytical generalized wiggler trajectories. In Sec.~\ref{acalc}, we derive exact expressions for the IC betatron wavelength and period as a function of the oscillation energy \(\Delta\gamma\). In Sec.~\ref{IC_und_match}, we join the results from the two previous sections, producing an expression for the plasma density as a function of \(K\) and \(\gamma_0\) to properly match IC and undulator trajectories, in terms of oscillation amplitude and wavelength. In Sec.~\ref{IC_spec}, we show the analytical expression for the critical frequency resolved over angle, showing unique features related to the high $(K/\gamma_0$) regime. In Sec.~\ref{results}, we show numerical checks of the presented analytical calculations, performed with the code \texttt{radyno}~\cite{radyno}. Finally, in Sec. ~\ref{Conclusion} we offer conclusions and outlook for future work.

\section{Generalized wiggler trajectories}\label{gen_wig_traj}

For low values of the $(K/\gamma$) parameter, undulator trajectories are well approximated by sinusoids. The oscillation amplitude $(x_0$) is given by the beam energy together with geometric and magnetic device properties as follows \cite{Ciocci2000}: 
\begin{equation}\label{dx_U}
    x_0=\frac{K}{\gamma k_U}        
\end{equation}
with $(\gamma$) the beam Lorentz factor, $(k_U$) the undulator wavenumber, and $(K$) the undulator strength. $(K$) is defined as the geometry-normalized magnetic field intensity:
\begin{equation}\label{K}
    K=\frac{B_0e}{m_eck_U}     
\end{equation}
where $(B_0$) represents the maximum amplitude of the ideally sinusoidal undulator magnetic field $(B(z)=B_0\cos{k_Uz}$). We are interested in comparing ion channel and undulator radiation in the high $(K/\gamma$) regime. Via numerical simulations, it was observed that, for $(K/\gamma\rightarrow1$), undulator trajectories are no longer sinusoidal and the oscillation amplitude diverges from that which is provided by Eq.~\ref{dx_U}. This is explained by the beam rigidity equation $(B\rho\approx\gamma m c/e$), where for $(K/\gamma=1$) the bending radius is equal to the inverse of the undulator wavenumber $(\rho=1/k_U$); consequently, the beam is bounded to the first device magnet and no undulation takes place. 

In contrast with an ideal undulator (with a purely sinusoidal magnetic field), an ideal ion channel device (with a purely linear focusing electric field) has a symmetry axis equal to the channel's central axis. To compare the radiation emitted from these two devices, particle trajectories should be matched to the greatest extent possible. Then, Eq.~\ref{dx_U} needs to be corrected to obtain the ion channel injection amplitude that meets the undulator trajectory amplitude \cite{Faure2010InjectionChannel}. Given the undulator field, the particle's bending radius can be expressed as a function of longitudinal position as follows:
\begin{equation}\label{b_radius}
    \rho=\frac{\gamma}{Kk_U\cos{k_Uz}}
\end{equation}
The differential of the two coordinates in the undulation plane will be simply given by:
\begin{equation}\label{diffs}
\begin{aligned} 
dz=\rho d\theta\cos{\theta} \\ 
dx=\rho d\theta\sin{\theta}
\end{aligned}
\end{equation}
with $(\theta$) the trajectory's angle with respect to the undulator's longitudinal axis. Substitution of Eq.~\ref{b_radius} in the first differential leads to the trajectory angle as a function of $(z$):
\begin{equation}\label{traj_ang}
    \theta=\arcsin\left(\frac{K}{\gamma}\sin k_Uz\right)
\end{equation}
The second differential in Eq.~\ref{diffs} can be now exploited to give an analytical \(x(z)\) expression:
\begin{equation}\label{wigg_traj}
    \begin{split}
    x(z)&=\frac{1}{2k_U}\Bigg[\ln(2) + 2\ln\left(1+\frac{K}{\gamma}\right) - \\-2\ln\Bigg(\sqrt{2}\frac{K}{\gamma}&\cos{k_Uz} + \sqrt{2 - \frac{K^2}{\gamma^2}(1-\cos{2k_Uz})}\Bigg) \Bigg]
    \end{split}
\end{equation}
This expression gives the correct undulator (wiggler) trajectories for $(K/\gamma$) in the range $([0,\,1)$), together with a much simpler expression for the trajectory oscillation amplitude:
\begin{equation}\label{dx_new}
    x_0=\frac{1}{k_U}\arctanh{\frac{K}{\gamma}}      
\end{equation}
Note that for \(K/\gamma\rightarrow 0\), Eq.~\ref{dx_new} tends to Eq.~\ref{dx_U}, while for $(K/\gamma\rightarrow 1$) the two expressions diverge. For $(K/\gamma \geq 1$) the expression is undefined, with a trajectory bending radius shorter than half an undulator period. This means that for such a field strength, beam particles will find themselves stuck at first undulator magnetic element.

\begin{figure}  
\includegraphics[width=\columnwidth]{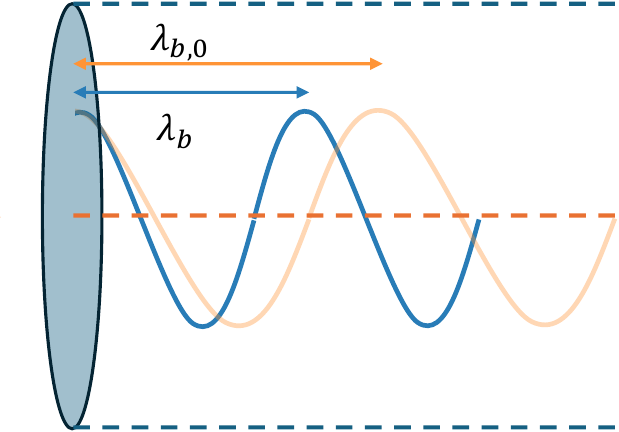}
\caption{Cartoon that shows wavelength modification in an IC. Even with the correct offset given by Eq.~\ref{dx_new}, the linear theory predicts $(\lambda_{b,0}$) but a shorter $(\lambda_b$) is actually observed. Two concurrent effects give the observed behavior. The real wavelength is shortened by particle velocity turning, becoming more relevant in the presence of high transverse momentum variations. It is then lengthened by particle energy variations, whose effect is to increase particle rigidity and raise the effective oscillation period.}
\label{fig:example}
\end{figure}

\section{Betatron Dynamics with Energy Oscillations}\label{acalc}

The betatron oscillation wavelength and period in the case of non-negligible energy oscillations may be found exactly in terms of elliptic integrals. The electron is injected into the uniform plasma with an initial longitudinal velocity $\beta_0$ at a transverse offset of $x_0$. The restoring force due to the plasma is $F=-kx$ with $k=\frac{e^2n_p}{2\epsilon_0}$ and $n_p$ the plasma density. By conserving longitudinal momentum we find the longitudinal and transverse velocities as a function of the Lorentz factor and the initial conditions,
\begin{equation}\label{bxbz}
    \begin{aligned}
    \beta_z&=\frac{\gamma_0}{\gamma(x)}\beta_0\\
    \beta_x^2&=1-\left(\frac{\gamma_0}{\gamma(x)}\right)^2,
    \end{aligned}
\end{equation}
with $\gamma_0$ the injection (initial) Lorentz factor.
From energy conservation, the total Lorentz factor $\gamma(x)$ may be found as a function of the transverse position,
\begin{equation}\label{gammax}
    \gamma(x)=\gamma_0 +\Delta\gamma\left[1-\left(\frac{x}{x_0}\right)^2\right],
\end{equation}
with the transverse injection offset $x_0$ and the oscillation energy magnitude
\begin{equation}\label{deltagamma}
\Delta\gamma=\frac{kx_0^2}{2mc^2}.
\end{equation}
Note that this magnitude is the full difference between maximum and minimum Lorentz factors, i.e. ${\gamma(x)\in[\gamma_0,\gamma_0+\Delta\gamma]}$.

\begin{figure}  
\includegraphics[width=\columnwidth]{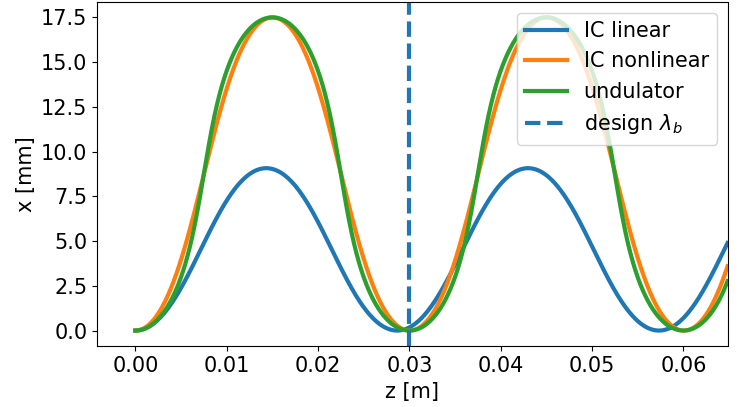}
\caption{Linear and nonlinear IC trajectories compared with undulator trajectories for $(K=95)$ and $(\gamma_0=100)$. In this regime, the linear theory (solid blue) fails to reproduce the design betatron wavelength. Additionally, the linearized injection offset $(x_0=K/\gamma_0 k_b)$ calculates the wrong oscillation amplitude. The presented theory (solid orange), with Eqs.~\ref{dx_new},~\ref{nps}, produces more closely matched results to the undulator trajectory (solid green) in terms of amplitude and wavelength. Note that the latter two trajectories still differ in core features like local curvature radius: this mismatch is the root cause of the radical difference in the devices' radiation.}
\label{traj_comp}
\end{figure}

The infinitesimal displacement along the longitudinal coordinate may be rewritten as ${dz=\frac{dz}{dt}\left(\frac{dx}{dt}\right)^{-1}dx=\frac{\beta_z}{\beta_x}dx}$. Thus, the length traveled in the longitudinal coordinate is found by integrating over a quarter period of the oscillation,

\begin{equation}
    \lambda_b=4\int_0^{x_0}dx\; \frac{\beta_z}{\beta_x},
\end{equation}
where $\lambda_b$, the betatron wavelength, is quadruple of the longitudinal length traveled in a quarter oscillation.
Substituting in Eqs. \ref{bxbz},

\begin{equation}
    \lambda_b=4\gamma_0\beta_0\int_0^{x_0} dx\; (\gamma^2(x)-\gamma_0^2)^{-\frac{1}{2}},
\end{equation}
and following the application of Eq. \ref{gammax} we attain the exact expression,

\begin{equation}\label{wl_correction}
\lambda_b=\frac{2}{\pi}\left(1+\frac{1}{2}\frac{\Delta\gamma}{\gamma_0}\right)^{-\frac{1}{2}}\mathcal{K}\left(\frac{\Delta\gamma}{2\gamma_0+\Delta\gamma}\right)\lambda_{b,0},
\end{equation}
with $\mathcal{K}$ the complete elliptic integral of the first kind. ${\lambda_{b,0}=2\pi c\beta_0\omega_{b,0}^{-1}}$ and $\omega_{b,0}=\sqrt{\gamma_0^{-1}\frac{k}{m}}$ are the betatron wavelength and frequency in the linear theory (${\Delta\gamma=0}$), respectively.
For weakly relativistic oscillations in the transverse coordinate, $\frac{\Delta\gamma}{\gamma_0}\ll1$, we have, to first order,

\begin{equation}\label{wl_correction_app} 
    \lambda_b=\lambda_{b,0}\left[1-\frac{1}{8}\frac{\Delta\gamma}{\gamma_0}+O\left(\frac{\Delta\gamma}{\gamma_0}\right)^2\right].
\end{equation}
The nonlinear theory then predicts that the betatron wavelength decreases for $\frac{\Delta\gamma}{\gamma_0}>0$, as depicted in Fig. \ref{fig:example} and seen in Fig. \ref{traj_comp}.

Using a similar expression for the oscillation period, ${T=\frac{4}{c}\int_0^{x_0} dx \beta_x^{-1}}$, yields

\begin{equation}\label{tl_correction_app}
\begin{alignedat}{3}
    &T_b&=4\omega_{b0}^{-1}&\bigg[
    2\left(1+\frac{1}{2}\frac{\Delta\gamma}{\gamma_0}\right)^{\frac{1}{2}}\mathcal{E}\left(\frac{\Delta\gamma}{2\gamma_0+\Delta\gamma}\right)\\
    &&&-\left(1+\frac{1}{2}\frac{\Delta\gamma}{\gamma_0}\right)^{-\frac{1}{2}}\mathcal{K}\left(\frac{\Delta\gamma}{2\gamma_0+\Delta\gamma}\right)
    \bigg]\\
    &&=2\pi\omega_{b0}^{-1}&\left[1+\frac{3}{8}\frac{\Delta\gamma}{\gamma_0}+O\left(\frac{\Delta\gamma}{\gamma_0}\right)^2\right],
\end{alignedat}
\end{equation}
with $\mathcal{E}$ the complete elliptic integral of the second kind. An increase in period for $\frac{\Delta\gamma}{\gamma_0}>0$ is thus observed.

\section{Matching Ion Channel and Undulator trajectories}\label{IC_und_match}
\begin{figure}  
\includegraphics[width=\columnwidth]{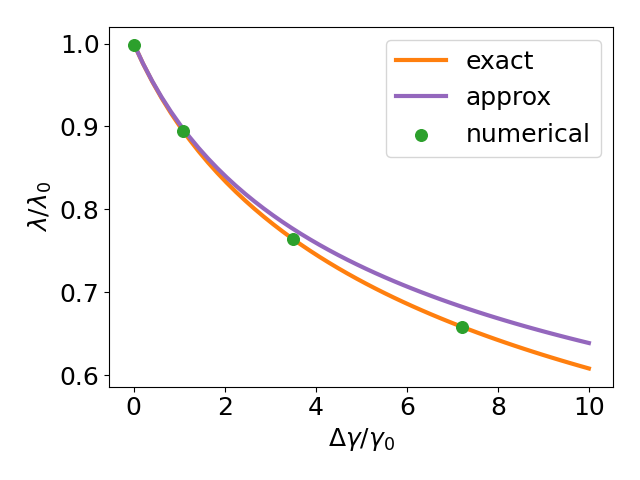}
\caption{Ratio between the real and linear IC betatron wavelength as a function of the relative energy oscillation amplitude. The solid orange line is computed with the analytical expression in Eq.~\ref{wl_correction}, which shows perfect agreement with numerical wavelength evaluation. The purple solid line is given by Eq. \ref{wl_correction_newapp}, which leads to an analytical plasma density expression as a function of particle energy and injection offset.}
\label{wavelen_ratio}
\end{figure}

With the corrections of Eqs.~\ref{dx_new} and \ref{wl_correction} it is now possible to compare the undulator and ion channel trajectories and radiation. A good approximation for Eq. \ref{wl_correction} is
\begin{equation}\label{wl_correction_newapp}
\lambda_b=\sqrt{\frac{8\pi^2\epsilon_0m_ec^2\gamma_0}{e^2n_p^2}}\left(\frac{1}{1+\frac{\Delta\gamma}{2\gamma_0}}\right)^{1/4}.
\end{equation}
It is found to differ from the exact Eq.~\ref{wl_correction} only up to $(\approx5\%$) for $(\Delta\gamma/\gamma_0\leq10)$, which is an extreme operational regime. Inverting Eq.~\ref{wl_correction_newapp}, a new expression for the plasma density is found as a function of the desired betatron wavelength and the relative energy oscillation:
\begin{equation}\label{np1}
    n_p \approx \frac{8\pi^2\epsilon_0m_ec^2\gamma_0}{e^2\lambda_b^2}\left(\frac{1}{1+\frac{\Delta\gamma}{2\gamma_0}}\right)^{1/2}.
\end{equation}

Inserting Eq.~\ref{deltagamma} in Eq.~\ref{np1}, the following equation is obtained and needs to be solved for $(\Delta\gamma)$:

\begin{equation}\label{GC}
    G^3+G^2-C^2=0
\end{equation}
with \(G=\Delta\gamma/2\gamma_0\), \(C=k_b^2x_0^2/4\) and $(x_0)$ again as the IC injection offset. Limiting solutions are found to be:

\begin{equation}\label{solutions}
   G=\begin{cases}
  C & \mbox{ if $ C\to0 $}\\
  C^{2/3}-\frac{1}{3}+\frac{1}{9C^{2/3}} & \mbox{ if $ C\to\infty $}
  \end{cases}
\end{equation}

The appropriate regime will depend on the injection condition and the desired wavelength. Substituting the wiggler injection offset found in Eq.~\ref{dx_new} gives $(C=\arctanh{(K/\gamma_0)}^2/4$) and an approximate solution for the energy oscillation magnitude as a function of $(K)$ and $(\gamma_0)$ is found. One of the two solutions in Eq.~\ref{solutions} may now be inserted back in the new plasma density expression in Eq.~\ref{np1} to provide the necessary plasma density such that the trajectories best match in both devices:

\begin{equation}\label{nps}
   n_p\approx\begin{cases}
  n_{p,0}\left(\frac{1}{1+C}\right)^{1/2} & \mbox{ if $ K/\gamma_0\to0 $}\\
  n_{p,0}\left(\frac{1}{2/3+C^{2/3}}\right)^{1/2} & \mbox{ if $ K/\gamma_0\to1 $}
  \end{cases}.
\end{equation}

With these expressions and Eq.~\ref{dx_new}, the same oscillation amplitude and wavelength are attained in the IC and undulator even for
$(K/\gamma_0\approx1)$. A proper comparison between radiation spectra can then be performed.

\section{Ion Channel spectrum for high 
$(K/\gamma)$}\label{IC_spec}

Low \(K\) regime spectral properties in ion channel devices have been widely explored \cite{Kostyukov2003, Rosmej2021}. The extreme energy oscillation related to the $(K/\gamma_0 \to 1$) regime leads to a radical change in spectrum energy and angular distribution. A theoretical prediction of these features will be presented. Considering ultra-relativistic electrons with $(\gamma_0\gg 1)$, the assumption $(K\gg 1)$ can be made as well, and the spectrum tends to a continuous synchrotron-like profile. Given the velocity components in Eq.~\ref{bxbz} and the relation between energy and transverse position Eq.~\ref{gammax}, the expression for the planar trajectory curvature radius \(\rho=|(1 + x'^2)^{3/2}/x''|\) may be found:

\begin{equation}\label{curvrad}
    \rho\approx\frac{m_e c^2}{k}\frac{\gamma^2}{\gamma_0}\frac{1}{\sqrt{2m_ec^2(\gamma_0-\gamma)/k + x_0^2}}
\end{equation}

\begin{figure}[h!]
\includegraphics[width=\columnwidth]{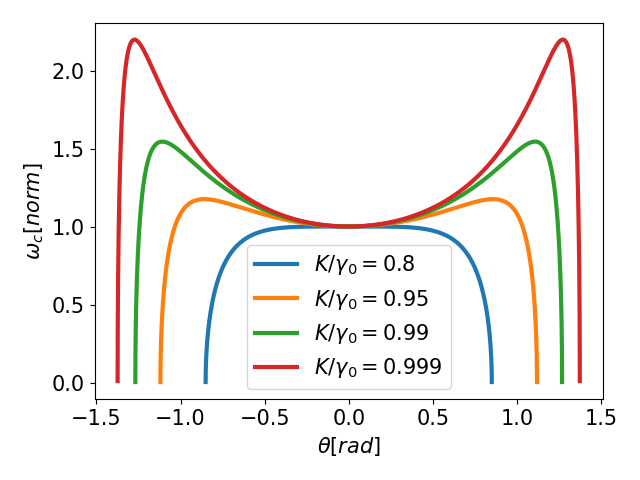}
\caption{Critical frequency versus angle with respect to the IC axis for several \(K/\gamma_0\) values, computed from Eq.~\ref{omegac}. As the ratio tends to 1, the well-known relativistic Doppler shift behavior is no longer dominant: the higher particle energy found near the axis results in intense emissions near the critical angle \(\pm\theta_c\) and an increased critical frequency.}
\label{critfreq}
\end{figure}

The particle's Lorentz factor may then be expressed as a function of the trajectory angle $(\theta)$ as

\begin{equation}\label{gammatheta}
    \gamma=\gamma_0\sqrt{1+\tan(\theta)^2}
\end{equation}

Substituting Eqs.~\ref{curvrad},~\ref{gammatheta} in the synchrotron critical frequency expression, \(\omega_c=3c\gamma^3/2\rho\), leads to the following result:

\begin{equation}\label{omegac}
    \begin{split}
        \omega_c=\frac{3}{2}\frac{k}{m_ec}\gamma_0^2\sec{(\theta)}\sqrt{x_0^2 - \frac{2\gamma_0m_ec^2}{k}(\sec{(\theta)}-1)}
    \end{split}.
\end{equation}

\begin{figure}  
\centering
    \subfigure{
    \hspace{-0.3cm}
        \begin{overpic}[width=0.8\columnwidth]{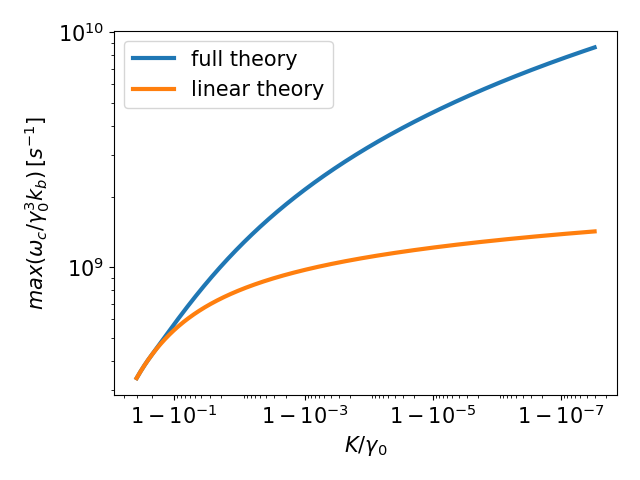}
            \put(88,17){\color{black}\textbf{(a)}}
        \end{overpic}
        \label{critfreq:a}
    }
    \subfigure{
        \hspace{0.0cm}
        \begin{overpic}[width=0.8\columnwidth]{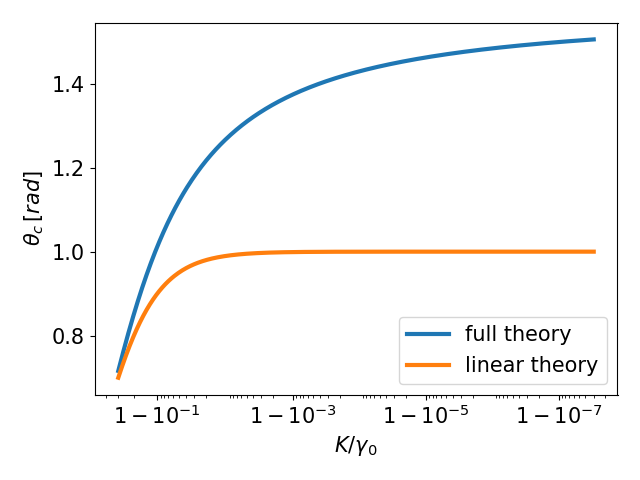}
            \put(18,65){\color{black}\textbf{(b)}}
        \end{overpic}
        \label{critfreq:b}
    }
    \caption{IC critical frequency analysis. (a) Maximal critical frequency versus $(K/\gamma_0)$. The linear theory (orange) predicts the highest critical frequency to be on axis. Much higher values may be found off axis as $(K/\gamma_0)$. This feature is shown by the present theory in solid blue curve. (b) Critical angle \(\theta_c\) versus \(K/\gamma_0\), a comparison between the linear theory and the present theory given by Eq.~\ref{thetac}. For \(K/\gamma_0\to1\), the peak intensity and critical frequency is found at \(\theta\approx\theta_c\), whose value differs substantially from the linear theory prediction as shown in the plot. Accounting for this difference is fundamental for proper experimental lines placing.}
    \label{critfreqan}
\end{figure}

The plasma density expression as found in Sec. ~\ref{IC_und_match} and the injection offset \(x_0\) from Eq.~\ref{dx_new} may be substituted in the previous expression, making it dependent on $(K)$ and $(\gamma_0)$. Figure~\ref{critfreq} shows the normalized $(\omega_c(\theta))$ for some $(K/\gamma_0)$ values. For lower $(K/\gamma_0)$ values, the well-known Doppler shifted angle dependence is observed, with the critical frequency monotonically decreasing as the observation point moves off axis (blue line). As $(K/\gamma_0\to 1)$, the trajectory's maximal slope and energy move towards divergence: energy grows faster than curvature radius and the combined effect is a stronger emission with higher frequencies at wide angles. The peak intensity is now found near the two critical angles $(\pm\theta_c)$. A comparison between previous and current theory for the peak photon energy as a function of $(K/\gamma_0)$ is presented in Fig.~\ref{critfreq:a}. The expression for the critical angle itself deviates from the well known $(\theta_c=K/\gamma_0)$, giving larger values as shown in Fig.~\ref{critfreq:b}:

\begin{equation}\label{thetac}
    \theta_c=\arccos\frac{1}{1+kx_0^2/2\gamma_0m_ec^2}
\end{equation}

This feature must be taken into account for proper placement of experimental lines at $(\pm\theta_c)$ with respect to the IC axis.

\subsection{Radiation fundamental frequency}

The fundamental radiation frequency is Doppler shifted from the trajectory frequency according to

\begin{equation}
 \omega_r=\omega\left(1-\frac{v_{p h s}}{c}\left(\hat{n} \cdot \hat{z}\right)\right)^{-1}   
\end{equation}

With $n_s$ a unit vector pointing to the observation direction and $v_{phs}$ the oscillation wave velocity. As $v_{p h s}$ decreases with added relativistic transverse motion the Doppler shift is diminished, resulting in a decreased radiation fundamental.

For $\hat{n}=\hat{z}$ (in front of the radiator) the resulting frequency is

\begin{equation}
    \omega_r=\left(1-\frac{v_{p h s}}{c}\right)^{-1} \omega
\end{equation}
With the first-order expansions of $v_{phs}=\frac{\lambda}{T}$ and $\omega=\frac{2\pi}{T}$ from Eqs. \ref{wl_correction_app} and \ref{tl_correction_app},

\begin{equation}
  \omega_r=\frac{\omega_{b0}}{1-\beta_0}\left[1-\frac{3+\beta_0}{8\left(1-\beta_0\right)} \frac{\Delta \gamma}{\gamma_0}+O\left(\frac{\Delta \gamma}{\gamma_0}\right)^2\right].
\end{equation}

And with $\beta_0 \approx 1,1-\beta_0 \approx \frac{1}{2} \gamma_0^{-2}$,

\begin{equation}\label{eq:plasmaUndulatorRadiationFrequency}
\omega_r\approx2 \gamma_0^2 \omega_{b0}\left(1-\Delta \gamma \gamma_0\right)
\end{equation}

The phase velocity must be relativistic for the Doppler shift to produce high frequencies. However, increased transverse motion rapidly decreases the phase velocity from $c$, quenching the Doppler effect. This consequence sets in rapidly, as the new parameter which must be ``small'' for this analysis is now $\Delta\gamma\gamma_0\ll1$. Ultimately, radiation spectra will have a much higher critical frequency than the radiation fundamental, drowning out the harmonic composition.

\section{Results}\label{results}

\begin{figure}[h!]  
\includegraphics[width=0.9\columnwidth]{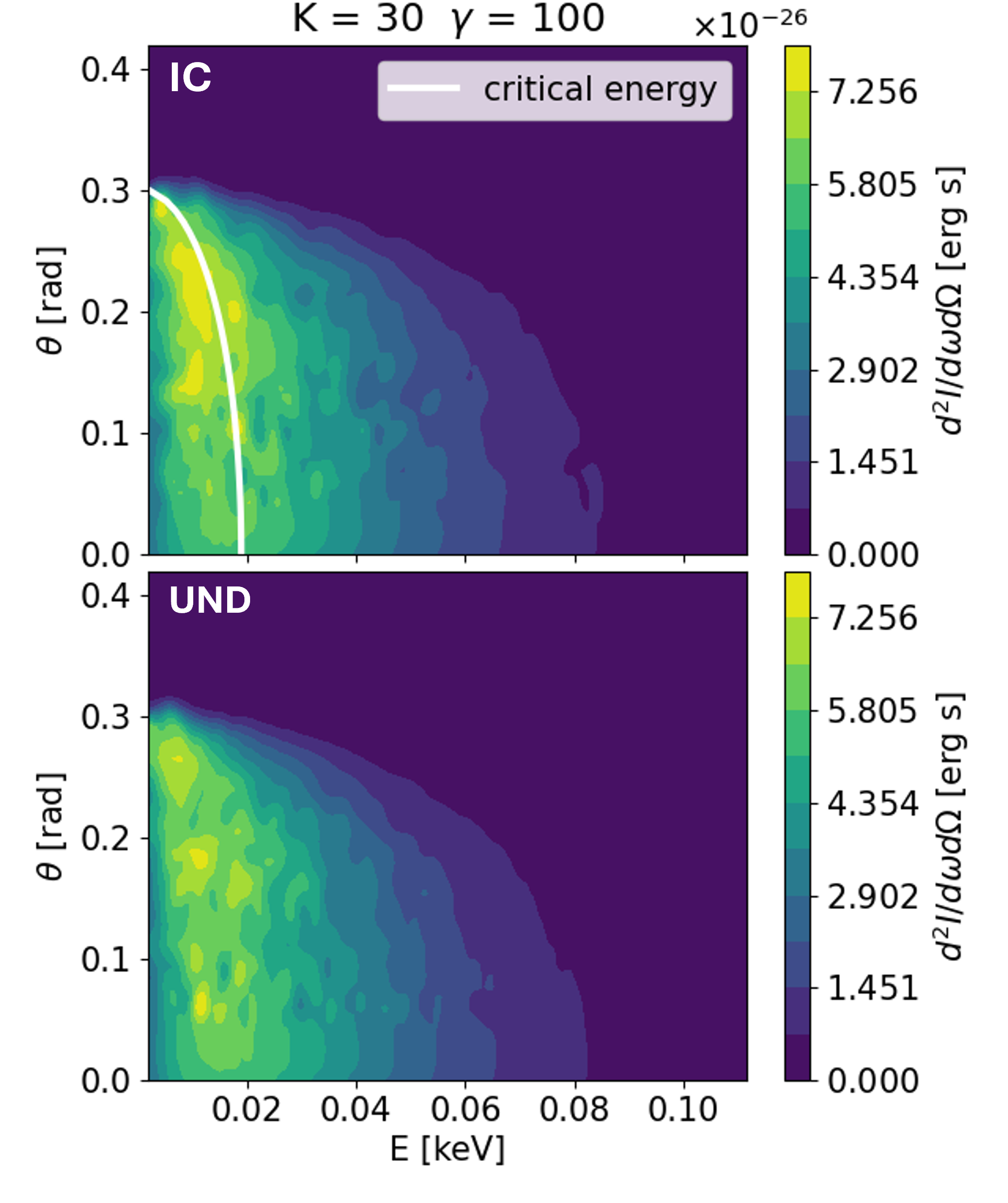} 
\caption{Ion channel VS undulator spectrum, \(K/\gamma_0=0.3.\)}
\label{fig:03}
\end{figure}

\begin{figure}[h!]
\includegraphics[width=0.9\columnwidth]{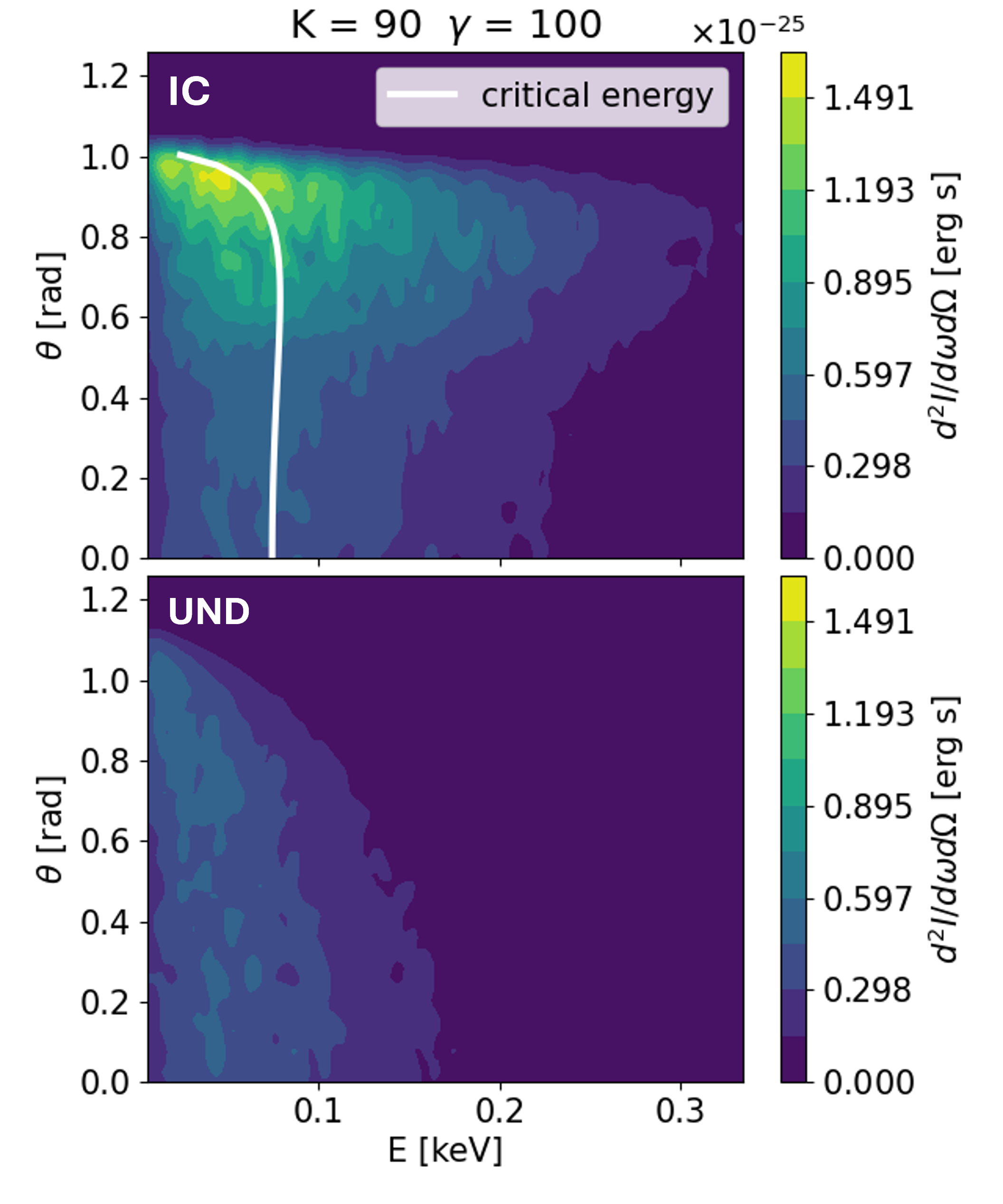} 
\caption{Ion channel VS undulator spectrum, \(K/\gamma_0=0.9.\)}
\label{fig:09}
\end{figure}

\begin{figure}[h!] 
\includegraphics[width=0.9\columnwidth]{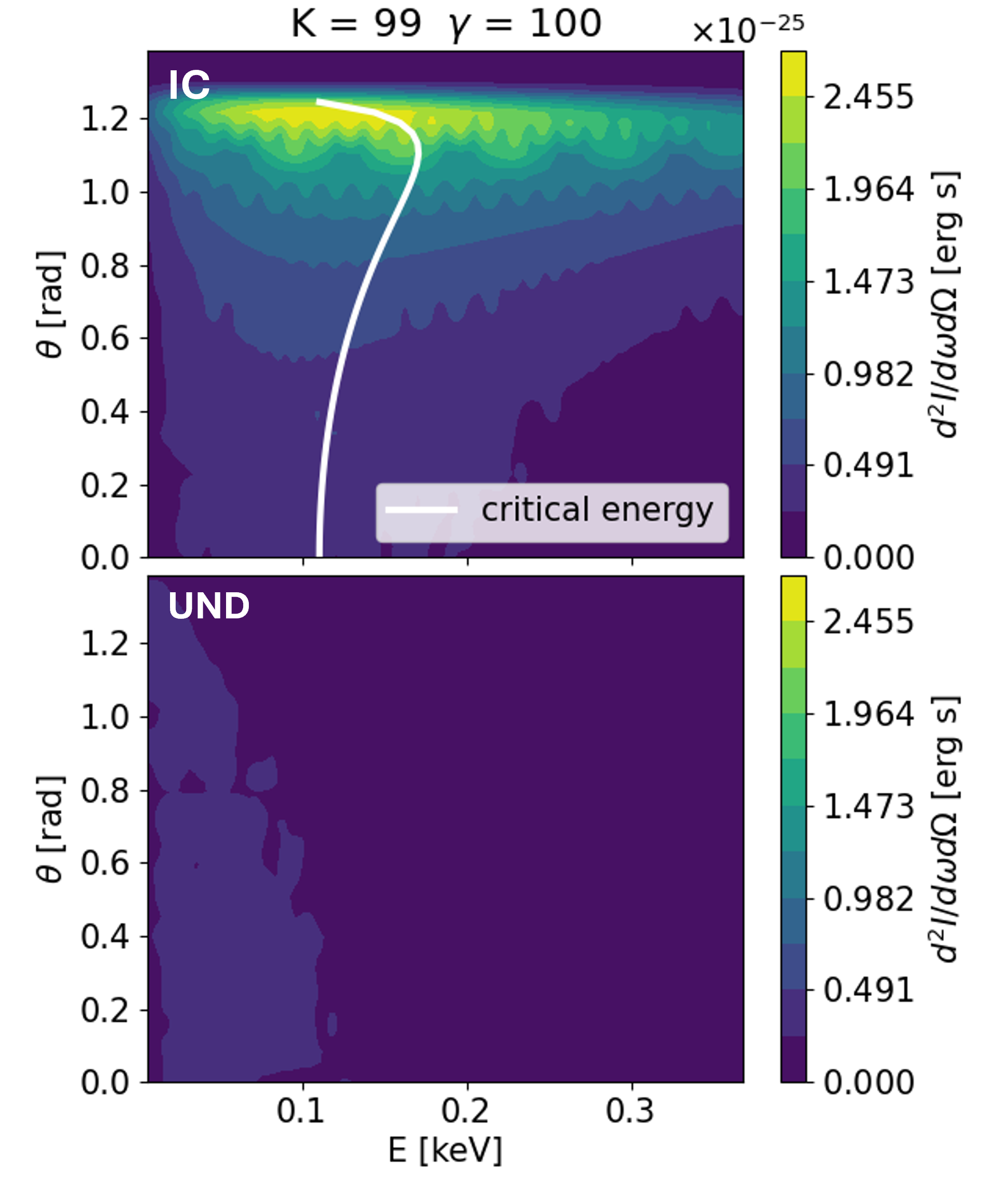} 
\caption{Ion channel VS undulator spectrum, \(K/\gamma_0=0.99.\)}
\label{fig:099}
\end{figure}

We perform a numerical comparison between IC and undulator trajectories and spectra with the \texttt{radyno} code~\cite{radyno}. Idealized IC and undulator fields have been implemented, as presented in Secs. ~\ref{gen_wig_traj},~\ref{acalc}. No wakefields or space charge effects are included, assuming highly relativistic particles. In Figs. ~\ref{fig:03},~\ref{fig:09},~\ref{fig:099} (\(K/\gamma_0 = 0.3,\,0.9,\,0.99\) and \(\gamma_0=100)\)), a comparison of single particle emission from an ion channel (IC, upper) and an undulator (UND, lower) is presented. The color charts show double differential intensity plotted versus photon energy (horizontal) and angle with respect to the device's axis (vertical). In the IC plots, the theoretical critical energy is plotted with Eq.~\ref{omegac} in solid white as a function of the angle. Color scales on the right of each plot have been set to the same extremes to highlight spectrum differences. In Fig. ~\ref{fig:03} we note that for \(K/\gamma_0=0.3\) there is little difference between the two spectra. They share the same critical energy profile. Given the particle energy variation of the order of \(\approx5\%\), the IC spectrum already features an intensity peak around \(\theta \approx0.2\,\text{[rad]}\). Moving to Fig. ~\ref{fig:09}, with \(K/\gamma_0=0.9\), a great difference is observed. IC peak intensity is now far from axis, around the critical angle, and exceeds undulator peak by more than a factor of 2 thanks to a \(\approx90\%\) energy gain. The critical frequency trend fits well with our analytical prediction. Figure~\ref{fig:099} shows an extreme regime where \(K/\gamma_0=0.99\) with an extraordinary energy gain of \(\approx240\%\). The relevant part of the IC spectrum now lies around the critical angle. The undulator spectrum is now five times less intense than the IC spectrum. The theoretical critical energy still follows the numerical spectrum.

\begin{figure}[h]  
\includegraphics[width=0.9\columnwidth]{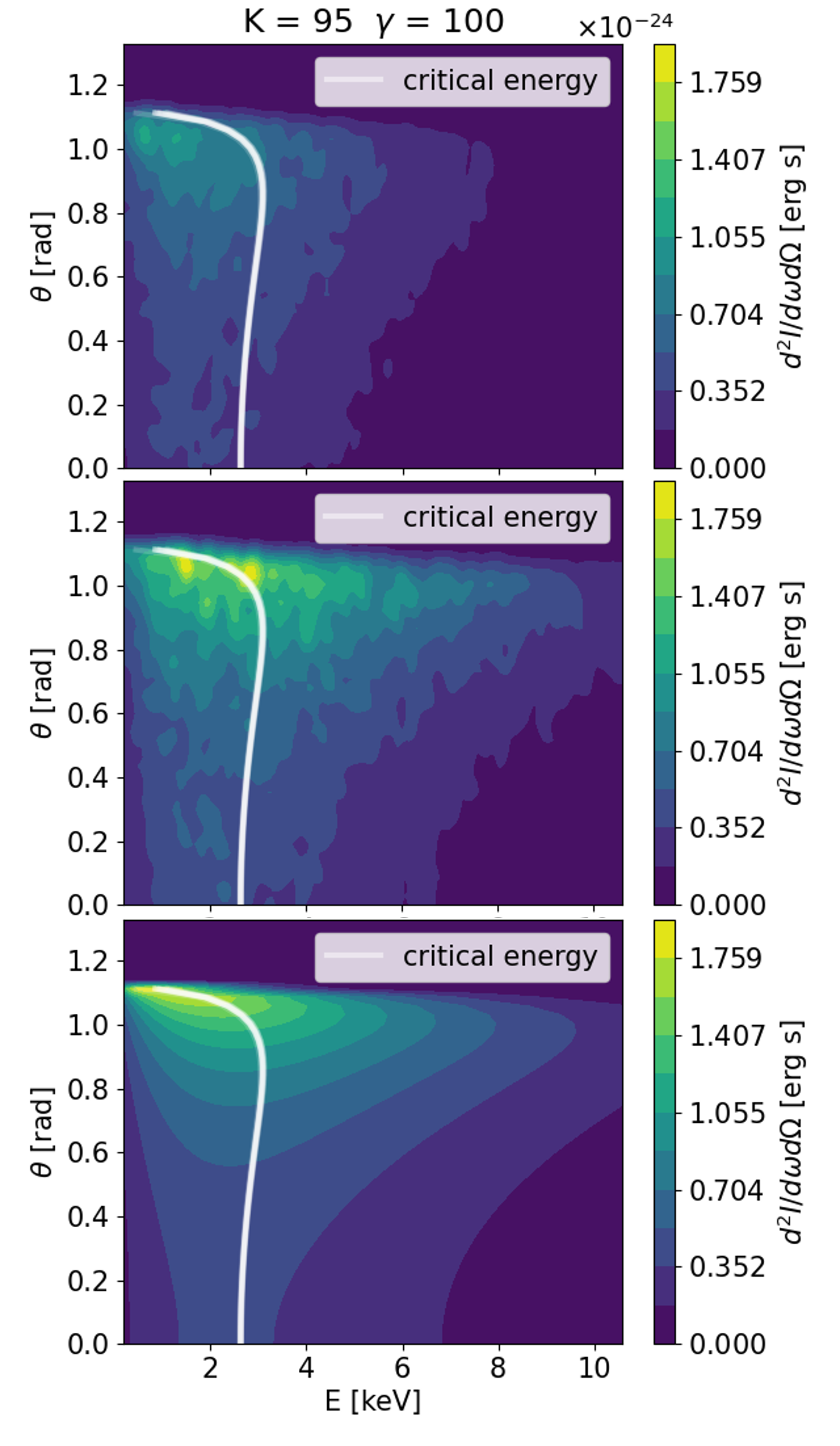} 
\caption{Full beam ion channel spectrum, \(K/\gamma_0 = 0.95\). Same emittance along two transverse planes (top), zero emittance along vertical plane (middle), analytical calculation (bottom).}
\label{beamfig}
\end{figure}

Up to now, only single particle behavior has been treated. To show a more realistic scenario, theoretical and numerical spectra from a full beam is now presented in Fig.~\ref{beamfig}. The injection energy is \(\approx50\text{MeV}\) and the design betatron wavelength is \(1\text{mm}\), then a desired \(K=95\) corresponds to a plasma density of \(1.7\times10^{17}\text{cm}^{-3}\). Such a setup requires an injection offset of \(\approx300\mu\text{m}\) and gives a \(130\%\) maximum energy gain. The selected beam transverse emittance is \(\epsilon_n=1\text{mm mrad}\), and no energy spread has been included.

\begin{figure}[h!]
\centering
    \subfigure{
        \hspace{-0.5cm}
        \begin{overpic}[width=0.8\columnwidth]{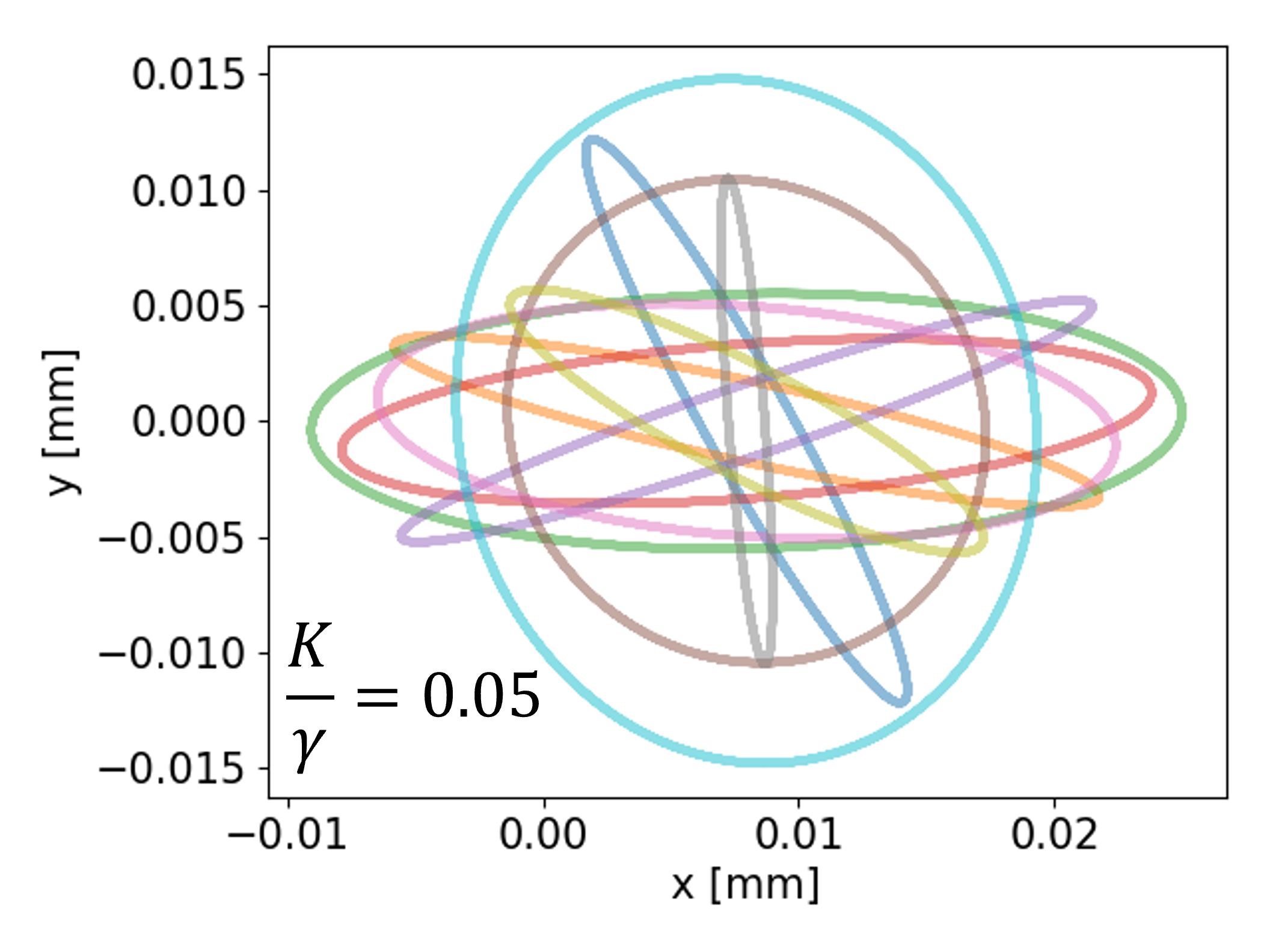}
            \put(88,17){\color{black}\textbf{(a)}}
        \end{overpic}
        \label{prec:a}
    }
    \subfigure{
        \begin{overpic}[width=0.8\columnwidth]{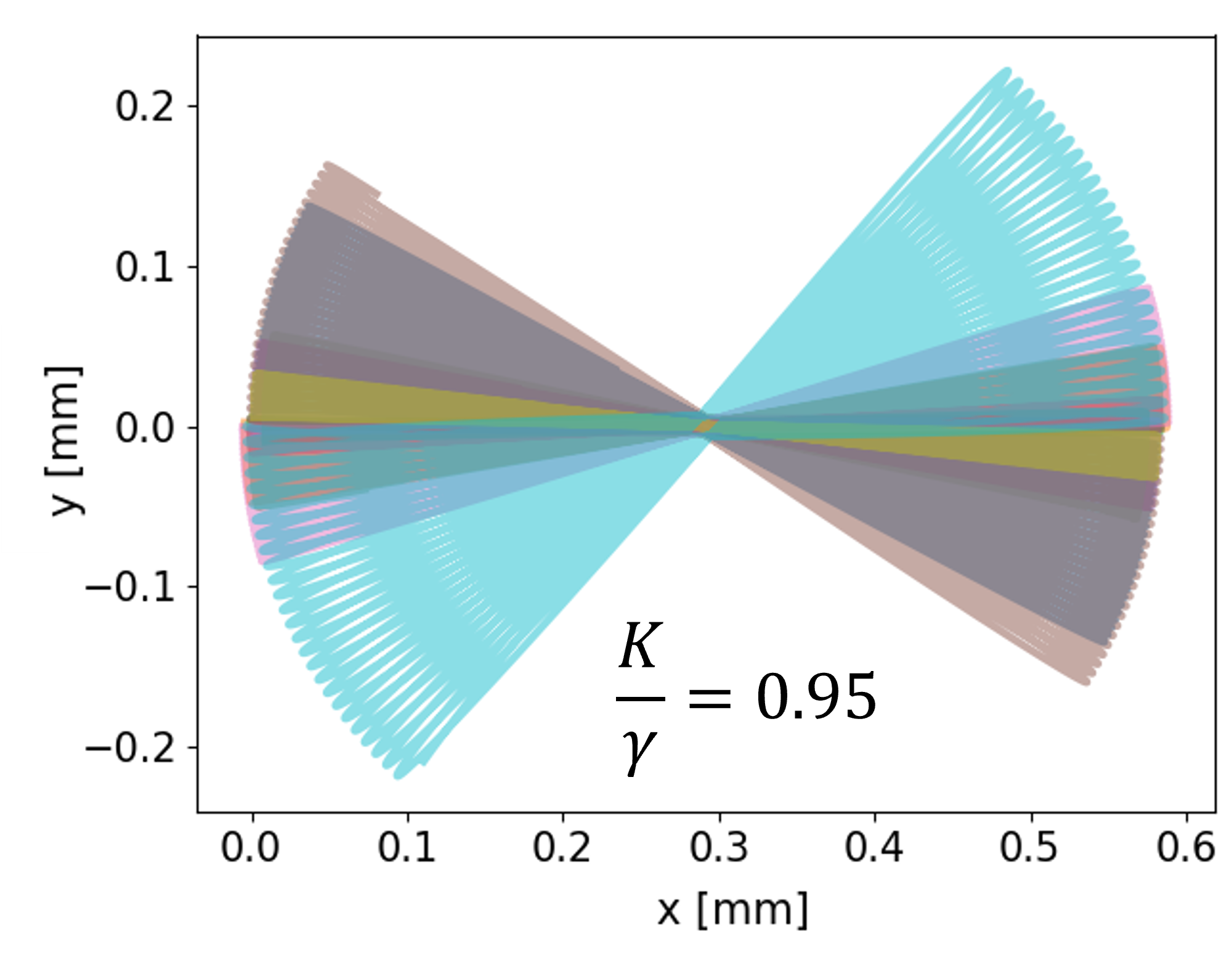}
            \put(88,17){\color{black}\textbf{(b)}}
        \end{overpic}
        \label{prec:b}
    }
    \caption{Precession phenomenon in transverse orbits. For low \(K/\gamma_0\) no orbit change is observed even after tens of periods (a). This is not the case when energy oscillation is high, as in case of \(H/\gamma_0=0.95\): here, the precession process is clear even for one period, with a fast transverse emittance growth (b).}
    \label{precession}
\end{figure}

The expected radiation spectrum shows an high intensity peak around \(2\text{keV}\) at \(\pm1.1\text{rad}\) (Fig.~\ref{beamfig} bottom). The numerical simulations correspond quite well with the theoretical predictions in case of zero vertical emittance (Fig.~\ref{beamfig} middle). A more realistic case of equal finite emittance along the two transverse axes shows a drop in intensity and a widening of the radiation spot in the vertical plane. This effect originates from a transverse orbit precession process induced by the high energy variation of beam particles. The difference in particle rigidity along the two transverse directions raises asymmetry in acceleration components, resulting in precession. No such behavior is observed for low energy variations. Transverse trajectories for low and high \(K/\gamma_0\) values are presented in Fig.~\ref{precession}, showing the appearance of precession in the latter case. The explanation and theory behind this phenomenon are discussed in detail in Appendices \ref{appA} and \ref{appB}.

\section{Conclusion} \label{Conclusion}

A fundamental comparison between ion channel and undulator single particle radiation emission at high \(K/\gamma_0\) was presented. The first issue was to properly match trajectories in the two devices. However, ion channel trajectories had smaller amplitudes and shorter wavelengths compared to those of an equivalent undulator following the existing linear theory. As we explored in this paper, this mismatch arose due to nonlinearities relevant for \(K/\gamma_0\to1\).

A proper undulator strength expression was calculated in Sec.~\ref{gen_wig_traj}, showing that \(K/\gamma_0=1\) defines the limit undulator strength \(K\); for higher values, no undulation takes place and particles are bound to the first magnetic element. The ion channel betatron oscillation wavelength shift was instead due to the intense energy variation due to transverse focusing with high \(K/\gamma_0\). In this regime, transverse energy gain can easily double the total energy from the particle's injection energy. A comprehensive expression for nonlinear betatron wavelength was calculated in Sec.~\ref{acalc}. Joining these results, a correction for plasma density was obtained as a function of desired trajectory features in Sec.~\ref{IC_und_match}. Moreover, a fully analytical expression for radiation critical frequency was calculated in Sec.~\ref{IC_spec}, showing a unique feature of high \(K/\gamma_0\) regime: particle energy grows as it moves towards device axis, leading to more intense and higher energy radiation emission at wide (\(\approx1\,\text{rad}\)) angles. This behavior gives two well-separated spots emission that may exceed undulator radiation energies and intensities. No particle-radiation interaction was taken into account since such wide radiation spread prevents superposition with trajectories.

Theoretical findings were compared with numerical radiation evaluation performed with \texttt{radyno} package. Full Gaussian beam simulations were performed as well, showing a spot deterioration with beam emittance. It was numerically shown that this effect is related to transverse plane orbit precession due to the intense transverse rigidity change during the betatron oscillation. Assuming oscillations take place in the horizontal plane, possible designs for such a radiation source should try to limit as much as possible the vertical beam emittance.

\section{Acknowledgement}
This work was performed with the support of the US Department of Energy, Division of High Energy Physics, under Contract No. DE-SC0009914, DE-SC0017648, NSF PHY-1549132 Center from Bright Beams, DARPA under Contract N.HR001120C007 and has been partially funded by the European Union — Next Generation EU.

\bibliography{main.bib}

\appendix
\section{Transverse orbit precession}\label{appA}

The transverse orbit precession phenomenon observed for the high \(K/\gamma_0\) regime will be further explored in the present section. As shown in Fig.~\ref{precession}, this effect is more observable for high energy variation. To exclude the possibility of numerical artifacts, a specific model for relativistic particle trajectories in the focusing electric field has been developed. For convenience, a coordinate change from Cartesian to cylindrical has been performed. Given the purely radial focusing force, conservation of angular momentum (\(L_z\)) in addition to longitudinal momentum (\(p_z\)) may be assumed. This gives the following velocity components:
\begin{equation}\label{betacomp}
    \begin{aligned}
        \beta_z = &\beta_{z,0}\frac{\gamma_0}{\gamma}\\
        \beta_\theta = &\beta_{\theta,0}\frac{\gamma_0 r_0}{\gamma r}\\
        \beta_r = &\sqrt{1 - \frac{1}{\gamma^2} - \beta_z^2 - \beta_\theta^2}
    \end{aligned}
\end{equation}
where \(\gamma\) is a function of radial coordinate \(r\) as defined in Eq.~\ref{gammax} after the substitution \(x\to r\). Note that this assumption requires \(\beta_{r,0}=0\), otherwise a new definition of \(\gamma(r)\) should be given. For the current purpose, this limitation won't be an issue at all and will only give restrictions on the possible particle starting conditions.

Differentiation of \(\beta_r\) leads to a second order differential equation for radial distance as a function of time:

\begin{equation}\label{rdd}
\begin{aligned}
\mathcal{A}(t) = &\beta_{\theta,0}^2\gamma_0^2r_0^2(3kr(t)^2 - \gamma_0mc^2 - kr_0^2)\\
\mathcal{B}(t) = &2k(1 + \beta_{z,0}^2\gamma_0^2)r(t)^4\\
\mathcal{C}(t) = &kr(t)^3 - (\gamma_0mc^2 + kr_0^2)r(t)\\
    r''(t) = &m^2c^6\frac{\mathcal{A}(t) + \mathcal{B}(t)}{\mathcal{C}(t)^3}
\end{aligned}
\end{equation}

\begin{figure}[t]
\centering
    \subfigure{
        \begin{overpic}[width=0.8\columnwidth]{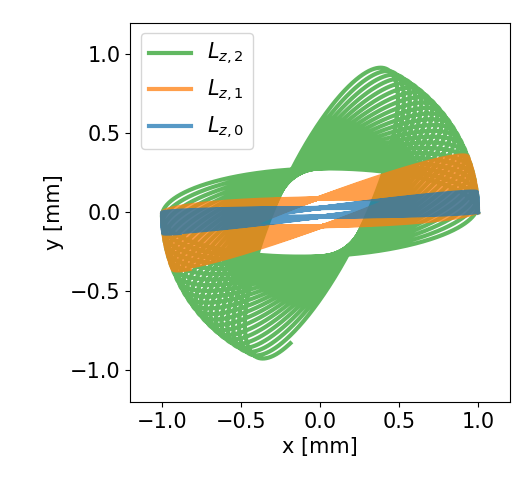}
            \put(88,20){\color{black}\textbf{(a)}}
        \end{overpic}
        \label{precsa:a}
    }
    \subfigure{
        \begin{overpic}[width=0.8\columnwidth]{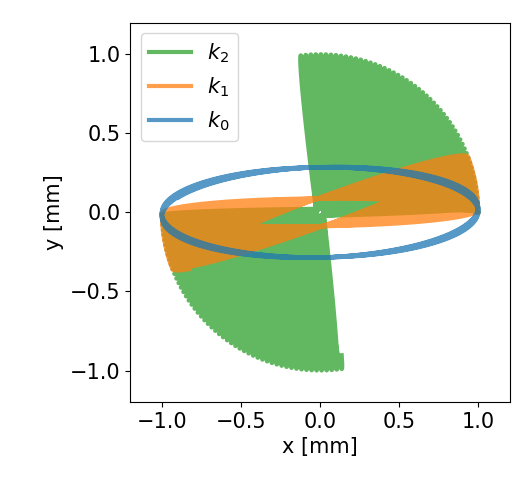}
            \put(88,20){\color{black}\textbf{(b)}}
        \end{overpic}
        \label{precsa:b}
    }
    \caption{Transverse orbit precession as computed from semi-analytical trajectory model. (a) Fixed focusing strength \(k\), growing particle's angular momentum from \(L_{z,0}\) to \(L_{z,2}\). (b) Fixed angular momentum, growing focusing strength from \(k_0\) to \(k_2\).}
    \label{precession_sa}
\end{figure}

This equation still needs to be numerically solved. Given the solution \(r(t)\), \(\theta(t)\) may be computed from \(\beta_\theta\) definition in Eq.~\ref{betacomp}. Despite the lack of a fully analytical solution, this model is based on strong assumptions over particle dynamics, ensuring more reliable results. As found in the purely numerical solutions, precession takes place when transverse speed components become relativistic. Given the longitudinal momentum conservation, this behavior is met for high particle energy variation regimes. In Fig.~\ref{precession_sa} some cases are presented as computed from the model, integrated for the same amount of time in each case. In Fig.~\ref{precsa:a} the focusing strength \(k=2\,10^{-6}\)~[N/m] and injection amplitude \(r_0=1\)~[mm] are kept constant, while angular momentum assumes the values \(L_{z,0}=3.8\,10^{-25}\), \(L_{z,1}=1.2\,10^{-24}\), \(L_{z,2}=3.8\,10^{-24}\)~[kg~\(m^2\)/s]. As angular momentum grows, the minor axis of elliptical trajectory sections grows as well: this is explained by the greater starting angular speed. Higher angular momentum gives faster precession rate too, as predicted in Eq. \ref{deltatheta_ke}. In Fig.~\ref{precsa:b}, angular momentum \(L_z=1.2\,10^{-24}\)~[kg~\(m^2\)/s] and injection amplitude \(r_0=1\)~[mm] are kept constant, while focusing strength gets the values \(k_0=2\,10^{-7}\), \(k_1=2\,10^{-6}\), \(k_2=2\,10^{-5}\)~[N/m]. As \(k\) grows, elliptical sections shrink because of the more intense focusing, while precession rate grows. Note that in the \(k_0\) case, i.e. weaker focusing and low energy variation, precession effect is nearly absent. 

\begin{figure}  
\includegraphics[width=0.99\columnwidth]{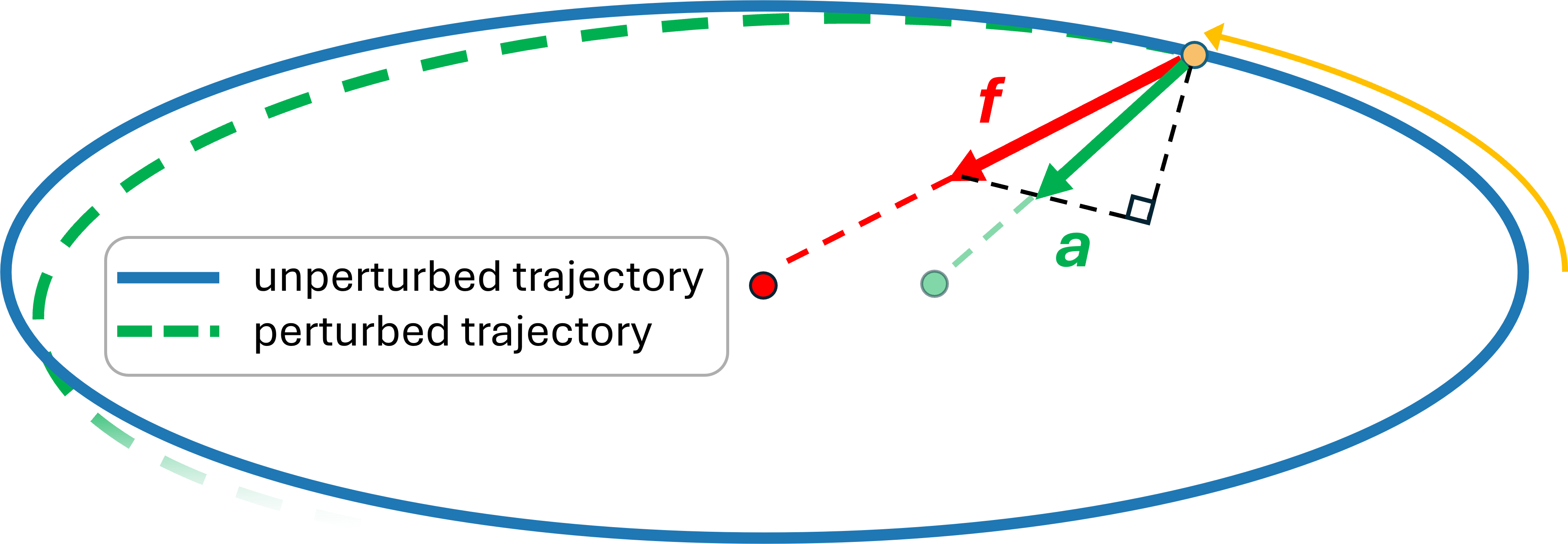} 
\caption{Cartoon scheme to show the relativistic acceleration effect that raises transverse orbit precession. Effective rotation center is shifted back and forth respect to device's axis, resulting in trajectory rotation.}
\label{precgraph}
\end{figure}

The arise of precession is explained thanks to relativistic acceleration expression
\begin{equation}\label{relacc}
    \bm{a} = \frac{\bm{f} - (\bm{f}\cdot\bm{\beta})\bm{\beta}}{m\gamma},
\end{equation}
where \(\bm{a}\), \(\bm{f}\), \(\bm{\beta}\) respectively stand for vector acceleration, force and normalized velocity. Referring to Fig.~\ref{precgraph}, when the \((x,y)\) velocity component gets relativistic, the dot product in Eq. \ref{relacc} is nonzero and the acceleration vector is shifted respect to the force vector. This effectively moves the instantaneous acceleration center (green dot in the figure) back and forth respect to device axis (red dot in the figure), steering the particle from the full blue to the dashed green trajectory.

\section{Precession angle per orbit}\label{appB}

To further quantify the orbital precession we aim to find the angle of precession per orbit. We start with a particle at position $r=r_0$ with angular velocity $\beta_{\theta,0}$, total gamma factor $\gamma_0$, zero radial velocity $\beta_{r,0}=0$, and a longitudinal velocity consistent with $\gamma_0$.

We will again use the integration technique of Sec. \ref{acalc}, this time for the excess angle drawn out in one full orbit,

\begin{equation}
    \Delta\theta=-2\pi + 4\int_{r_{min}}^{r_0}\;d\theta
\end{equation}

And we refactor the integrand with ${ d\theta=\frac{d\theta}{dt}\left(\frac{dr}{dt}\right)^{-1}dr }$ ${=\frac{1}{r}\beta_\theta\beta_r^{-1}dr}$. However, we must find $r_{min}$. Following Equations \ref{betacomp}, we express $\beta_r^2\gamma^2\left(\frac{r}{r_0}\right)^2$ as a cubic polynomial and divide by the known root ${x=\left(\frac{r}{r_0}\right)^2=1}$, resulting in a quadratic for the minimum position

\begin{equation}
    \left(\frac{r_{min}}{r_0}\right)^2=x_m=
    \left(\frac{\gamma_0}{\Delta\gamma}+\frac{1}{2}\right)\left[1-\sqrt{
    1-\frac{\beta_{\theta,0}^2}{\left(1+\frac{1}{2}\frac{\Delta\gamma}{\gamma_0}\right)^2}
    }\right].
\end{equation}

Note that, now that the particle does not reach the ion channel axis, $\Delta\gamma$ is better understood as the initial potential energy relative to the axis center, as opposed to the energy oscillation magnitude. For $\beta_{\theta,0}\ll1$ we have $x_m\approx\frac{\beta_{\theta,0}^2}{\frac{\Delta\gamma}{\gamma_0}\left(2+\frac{\Delta\gamma}{\gamma_0}\right)}$. Knowing these two roots allows us to factor $\beta_r$

\begin{equation}
    \beta_r^2=
    \left(\frac{\Delta\gamma}{\gamma_0}\right)^2
    \left(\frac{\beta_\theta}{\beta_{\theta,0}}\right)^2
    \left(2\frac{\gamma_0}{\Delta\gamma}+1-x_m-x\right)
    (x-x_m)(1-x).
\end{equation}

This simplifies the integral somewhat. Following Equation 3.137.3 in Ref. \cite{table_series_integrals} and making appropriate adjustments for our integral we ultimately arrive at the exact expression for the precession angle per orbit

\begin{equation}
    \begin{aligned}
        \Delta\theta=-2\pi&+
        4\frac{\gamma_0}{\Delta\gamma}\beta_{\theta,0}
        \frac{1}{x_m\sqrt{2\frac{\gamma_0}{\Delta\gamma}+1-2x_m}}\\
        &\times\Pi_e\left(
        1-x_m^{-1};
        \frac{1-x_m}{2\frac{\gamma_0}{\Delta\gamma}+1-2x_m}
        \right),
    \end{aligned}
\end{equation}

with $\Pi_e(n;k)$ the complete elliptic integral of the third kind, $n$ the elliptic characteristic, and $k$ the elliptic modulus. Note that our notation here is slightly different from that in Ref. \cite{table_series_integrals} where the second parameter is $q=\sqrt{k}$.

With the assumption of small $x_m\approx\frac{\beta_{\theta,0}^2}{\frac{\Delta\gamma}{\gamma_0}\left(2+\frac{\Delta\gamma}{\gamma_0}\right)}$ (implying strong focusing and/or low initial angular velocity) we have a more readily interpreted expression

\begin{equation} \label{deltatheta_ke}
    \begin{aligned}
        \Delta\theta&=4\frac{\gamma_0}{\Delta\gamma}\left(1+2\frac{\gamma_0}{\Delta\gamma}\right)^{-1/2}\\
        &\times\left[
        \mathcal{K}\left(\frac{1}{1+2\frac{\gamma_0}{\Delta\gamma}}\right)-
        \mathcal{E}\left(\frac{1}{1+2\frac{\gamma_0}{\Delta\gamma}}\right)
        \right]\beta_{\theta,0}\\
        &+O\left(\frac{\gamma_0}{\Delta\gamma}\beta_{\theta,0}^3\right)
    \end{aligned}
\end{equation}

\begin{figure}[t]  
\includegraphics[width=0.9\columnwidth]{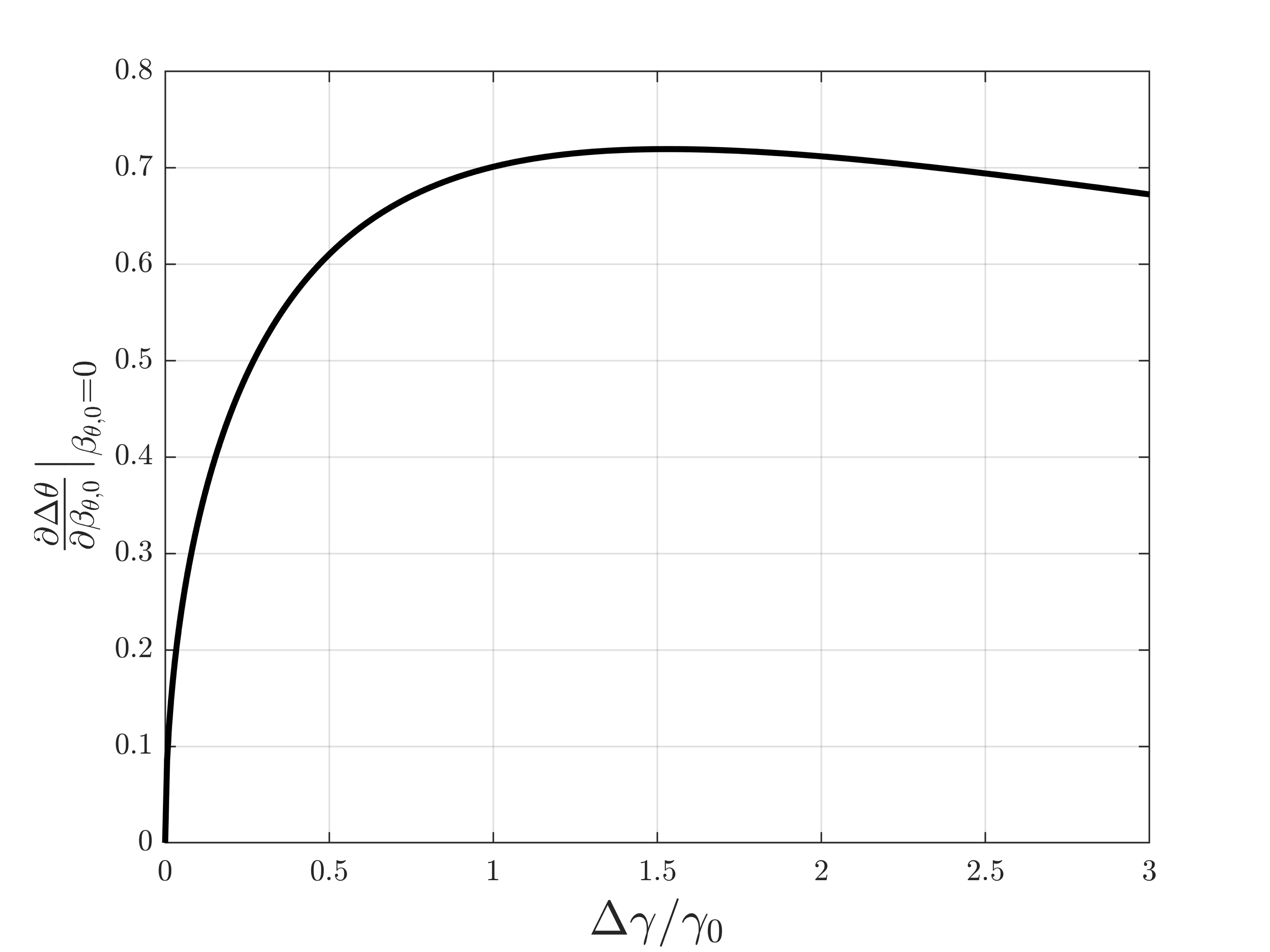} 
\caption{First-order coefficient for the precession angle per orbit. There is no precession without relativistic transverse motion.}
\label{fig:precession_angle_per_orbit}
\end{figure}

From here it is clear that the precession angle per orbit is approximately linear with the initial angular velocity. If this initial velocity is due to emittance then this approximation holds quite well. The dependence on $\frac{\Delta\gamma}{\gamma_0}$ is shown in Fig. \ref{fig:precession_angle_per_orbit}.

Further, for weakly relativistic transverse oscillations, $\frac{\Delta\gamma}{\gamma_0}\ll1$,

\begin{equation}
    \Delta\theta=\frac{\pi}{2\sqrt{2}}\sqrt{\frac{\Delta\gamma}{\gamma_0}}\beta_{\theta,0}
    +O\left(\left(\frac{\Delta\gamma}{\gamma_0}\right)^{\frac{3}{2}}, \beta_{\theta,0}^3\right).
\end{equation}

\end{document}